\documentclass{article}
\usepackage{abstract}
\usepackage{spconf,amsmath,graphicx,hyperref}
\usepackage{algorithm}
\usepackage{algpseudocode}
\usepackage{tabularx}
\usepackage{booktabs}
\usepackage{xcolor}
\usepackage{multirow}
\usepackage{colortbl}
\usepackage{graphicx}
\usepackage{subcaption}
\graphicspath{{figures/}}

\title{Marco-ASR: A Principled and Metric-Driven Framework for Fine-Tuning Large-Scale ASR Models for Domain Adaptation}

\name{\begin{tabular}[t]{c}
      Xuanfan Ni, Fei Yang, Fengping Tian, Qingjuan Li, Chenyang Lyu\textsuperscript{*}\thanks{* Corresponding Author.}, Yichao Du, \\ \it Longyue Wang, Weihua Luo, Kaifu Zhang 
      \end{tabular}}
\address{Alibaba International Digital Commerce}

\begin{document}
\ninept
\maketitle
\begin{abstract}

Automatic Speech Recognition (ASR) models have achieved remarkable accuracy in general settings, yet their performance often degrades in domain-specific applications due to data mismatch and linguistic variability. This challenge is amplified for modern Large Language Model (LLM)-based ASR systems, whose massive scale and complex training dynamics make effective fine-tuning non-trivial. To address this gap, this paper proposes a principled and metric-driven fine-tuning framework for adapting both traditional and LLM-based ASR models to specialized domains. The framework emphasizes learning rate optimization based on performance metrics, combined with domain-specific data transformation and augmentation. We empirically evaluate our framework on state-of-the-art models—including Whisper, Whisper-Turbo, and Qwen2-Audio—across multi-domain, multilingual, and multi-length datasets. Our results not only validate the proposed framework but also establish practical protocols for improving domain-specific ASR performance while preventing overfitting.

\end{abstract}


\section{Introduction}
\begin{figure}[t]
   \centering
   \includegraphics[width=\linewidth]{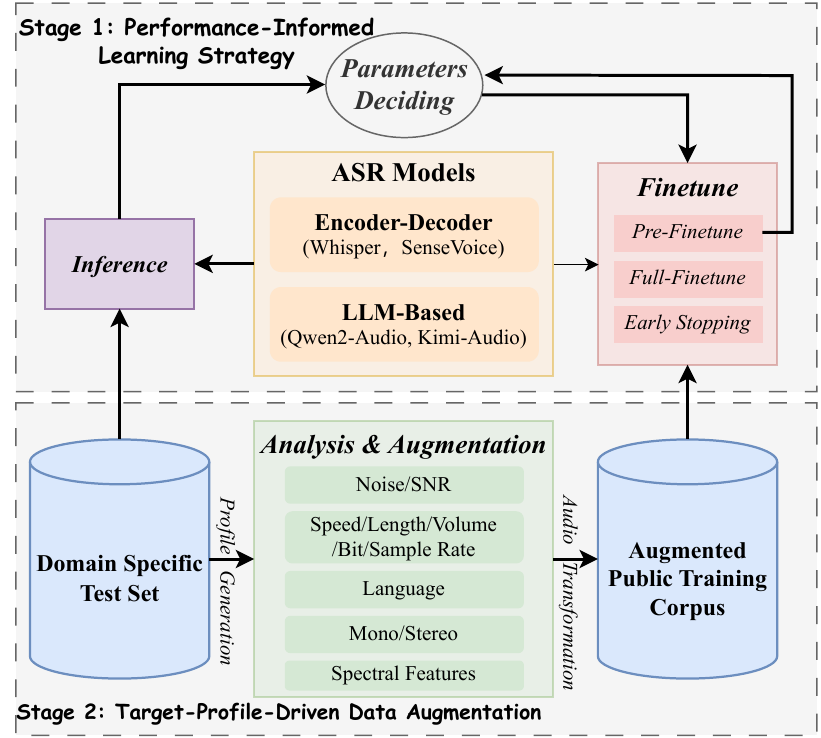}
        \caption{An overview of our framework, which consists of: (1) hyper-parameter optimization based on model architecture and performance, and (2) domain-specific data analysis and augmentation.}
    \label{workflow}
\end{figure}

Automatic Speech Recognition (ASR) systems, powered by large-scale pre-trained models like Whisper \cite{radford2022robust_whisper}, have become pervasive in various applications. More recently, the integration of Large Language Models (LLMs) has given rise to a new generation of powerful audio-foundation models, such as Qwen2-Audio \cite{chu2024qwen2audiotechnicalreport} and Kimi-Audio \cite{kimiteam2025kimiaudiotechnicalreport}, which exhibit even stronger capabilities in understanding and transcribing speech. However, despite their impressive performance on general-purpose benchmarks, these off-the-shelf models often falter when deployed in specialized, or out-of-domain (OOD), scenarios like medical transcription, legal dictation, or financial calls. This performance drop is primarily due to distribution shifts in acoustic conditions, speaker characteristics, and, most critically, domain-specific terminology.

While fine-tuning is a conventional approach for domain adaptation, applying it to modern LLM-based ASR models presents unique and significant challenges. Their vast parameter spaces and intricate architectures make them highly sensitive to training configurations. For instance, selecting an appropriate learning rate is not merely a matter of optimization but is critical to prevent catastrophic forgetting or training instability. Furthermore, our empirical findings reveal that traditional indicators of training progress, such as a decreasing loss, do not reliably correlate with improvements in transcription accuracy (i.e., Word Error Rate, WER~\cite{DBLP:conf/interspeech/MorrisMG04}), making the fine-tuning process opaque and difficult to control. These complexities necessitate a more structured and principled approach to adaptation.

To address these issues, this paper introduces a principled and metric-driven fine-tuning framework designed to be applicable to both traditional encoder-decoder and newer LLM-based ASR models. Our framework provides a systematic guide through the critical stages of domain adaptation: (1) principled hyperparameter selection, with a focus on identifying optimal learning rates through pre-finetuning trials and continuous model performance analysis using validation metrics and (2) rigorous data filtering and augmentation to prepare high-quality, domain-relevant training sets.

Through extensive experiments, we first demonstrate the extent of the OOD problem by showing how leading models underperform on various domain-specific and low-resource language datasets. We then apply our proposed framework to systematically fine-tune these models, presenting detailed results that highlight the impact of learning rate choices and the effectiveness of data augmentation. Our analysis provides practical insights and establishes a clear protocol for practitioners to effectively adapt large-scale ASR models for their specific needs, transforming them from general-purpose tools into highly accurate, domain-specialized solutions.

\section{A Principled, Metric-Driven Domain Adaptation Framework}
We formalize our framework for fine-tuning ASR models on domain adaptation, as shown in Fig~\ref{workflow}.
\subsection{Performance-Informed Learning Strategy}
The fine-tuning framework is initiated by quantifying a baseline performance metric on the target domains by calculating the WER on a subset of the target dataset (e.g., 200 instances). This baseline metric, in tandem with the model's architectural paradigm, is a critical determinant for selecting the optimal initial learning rate ($\eta$). 

\subsubsection{Encoder-Decoder ASR Models}
Fine-tuning these models requires a balanced learning rate. A low rate is stable but slow, while a high rate can destabilize training and cause the WER to fluctuate. To address this trade-off, we employ a dynamic learning rate strategy based on discrete training cycles. Our strategy partitions the fine-tuning process into cycles, where each cycle $j$ uses a constant learning rate, $\eta_j$. Throughout cycle $j$, the model is trained for a fixed number of steps, with periodic evaluations on the validation set. This process generates a list of WER measurements within that single cycle. We then quantify the training stability by calculating the standard deviation of this list of WERs, denoted as $\sigma_{\text{WER}}(j)$.

The learning rate for the subsequent cycle, $\eta_{j+1}$, is adapted based on training stability. A high $\sigma_{\text{WER}}(j)$ indicates instability, prompting a smaller learning rate, while a low value signals stable convergence, allowing for a larger learning rate. This is formalized by the update rule:
\begin{equation}
    \eta_{j+1} = \eta_{\max} - (\eta_{\max} - \eta_{\min}) \cdot \text{clip}\left(\frac{\sigma_{\text{WER}}(j)}{\sigma_{\text{ref}}}, 0, 1\right)
    \label{eq:lr_adaptive_cyclic}
\end{equation}
where $[\eta_{\min}, \eta_{\max}]$ is the allowed learning rate range, and $\sigma_{\text{ref}}$ is a reference standard deviation.

\subsubsection{LLM-Based ASR Models}
For LLM-based ASR models like \textit{Qwen2-Audio} and \textit{Kimi-Audio}, we conceptualize the initial baseline $\text{WER}_0$, as a quantitative proxy for the domain mismatch between the source (pre-training) and target datasets. The learning rate, $\eta_{\text{llm}}$, is therefore dynamically calibrated as a function of this domain gap. A significant mismatch, manifested as a high $\text{WER}_0$, necessitates a larger learning rate for accelerated convergence and substantial parameter adaptation. Conversely, a minimal domain gap (low $\text{WER}_0$) warrants a smaller rate for fine-grained refinement. This dynamic scaling is formalized as:
\begin{equation}
    \eta_{\text{llm}} = \eta'_{\min} + (\eta'_{\max} - \eta'_{\min}) \cdot \text{clip}(\text{WER}_0, 0, 100)/100
    \label{eq:lr_llm}
\end{equation}
where $\eta'_{\min}$ (e.g., $1 \times 10^{-6}$) and $\eta'_{\max}$ (e.g., $1 \times 10^{-4}$) define the learning rate boundaries for this model family. The $\text{clip}(\cdot)$ function constrains the $\text{WER}_0$ value to the interval $[0, 1]$, allowing the learning rate to scale linearly with the perceived domain gap.

We fine-tune the ASR model starting with an empirically-derived initial learning rate. During this process, the learning rate is adjusted based on the WER in accordance with the formulas above, and training is halted once the WER converges and shows no significant improvement. The empirical validation and comparative analysis of these learning rate strategies are detailed in Section 4.2.

\subsection{Target-Profile-Driven Data Augmentation}
To bridge the domain gap between public training corpora and domain datasets, we employ a profile-driven data augmentation strategy. This two-phase strategy first analyzes audio from the target domain to build a statistical profile, which then guides the augmentation of clean source data to realistically emulate the target's characteristics.
    
In the analysis phase, we generate a quantitative profile of the target audio domain from a representative corpus. This process involves analyzing each audio file to extract a set of technical attributes, namely \textbf{sample rate}, \textbf{bit depth}, and \textbf{channel count}. Concurrently, we compute key acoustic descriptors, including the Signal-to-Noise Ratio (\textbf{SNR}), integrated loudness (\textbf{LUFS}), the mean \textbf{Spectral Centroid} for perceptual brightness, and the \textbf{Spectral Rolloff} for spectral shape. The final profile summarizes the dataset by defining a characteristic range of values for each of these metrics.

The synthesis phase transforms a clean source audio signal $x(t)$ using the generated profile. First, the signal is resampled to a target sample rate and bit depth randomly chosen from the profile's distributions. Subsequently, a series of augmentations are applied. Reverberation is introduced via convolution with a randomly selected Room Impulse Response (RIR), with the application probability being inversely related to the profile's mean SNR. Background noise is added, with the target SNR of the resulting mixture being sampled from the range $[\text{SNR}_{\text{min}}, \text{SNR}_{\text{max}}]$ defined in the profile. The audio's loudness is then normalized by applying a gain factor to match a target LUFS value drawn from a normal distribution $\mathcal{N}(\mu_{\text{LUFS}}, \sigma_{\text{LUFS}}^2)$ derived from the profile's statistics. Finally, to emulate channel effects, spectral shaping is performed using a low-pass or high-pass filter. The choice of filter and its cutoff frequency are guided by the profile's mean spectral rolloff, thereby simulating conditions such as muffled or telephonic audio.

The entire augmentation process, which transforms a clean signal $x(t)$ into its augmented version $x_{\text{aug}}(t)$, can be formally expressed as a chain of operators:
\begin{equation}
x_{\text{aug}}(t) = (F_{\text{filt}} \circ F_{\text{lufs}} \circ F_{\text{noise}} \circ F_{\text{rev}} \circ R)(x(t))
\label{eq:augmentation_chain}
\end{equation}
where $\circ$ denotes function composition. The operator chain begins with resampling ($R$) to a target sample rate and bit depth. Next, $F_{\text{rev}}$ probabilistically applies reverberation ($*h_{\text{rir}}$) with a probability $p_{\text{rev}} \propto 1/\mu_{\text{SNR}}$. Subsequently, $F_{\text{noise}}$ adds noise to achieve an SNR drawn from $U(\text{SNR}_{\text{min}}, \text{SNR}_{\text{max}})$, and $F_{\text{lufs}}$ normalizes the loudness to a target LUFS from $\mathcal{N}(\mu_{\text{LUFS}}, \sigma_{\text{LUFS}}^2)$. The final step, $F_{\text{filt}}$, applies a spectral filter ($*h_{\text{filt}}$) guided by the mean spectral rolloff $\mu_{\text{rolloff}}$.

\begin{table*}[!t]
\centering
\resizebox{0.8\textwidth}{!}{
\begin{tabular}{lllllllllll}
\toprule[1.3pt]
\textbf{Models} & \textbf{\#Param} & \textbf{Zh} & \textbf{En}    & \textbf{Es} &
\textbf{Ja}     & \textbf{Th}    & \textbf{Tr}     & \textbf{Vi}     & \textbf{Ko}    & \textbf{Id}    \\ \hline
\multicolumn{11}{c}{\cellcolor{gray!40}\textit{\textbf{Encoder-Decoder}}} \\
Whisper-Large-V3   & 1.5B           & 24.77 & 10.26 & 9.14& 29.35  & 21.92 & 17.20  & 31.46  & 9.50  & 13.67 \\ 
Whisper-Large-V3-Turbo  & 0.8B     & 24.80 & 9.90  &11.28& 46.55  & 41.62 & 29.40  & 31.34  & 9.32  & 21.48 \\
SenseVoice     & 0.2B         & 15.11 & 12.73 &  -- --&37.13  &  -- --   & -- --     &  -- --    & 32.41 &  -- --   \\
SeamlessM4T      & 2.3B        & 12.34 & 4.05&4.92  & 237.60 & 4.29  & 11.71  & 18.74  & 32.74 & 6.27  \\
FireRedASR-AED & 1.1B & 15.11 & 12.73 & -- -- & 37.13 &  -- -- &  -- -- &  -- -- & 32.41 &  -- -- \\ \hline
\multicolumn{11}{c}{\cellcolor{gray!40}\textit{\textbf{LLM-Based}}} \\
Qwen2-Audio-Base   & 8.4B           & 20.51 & 5.47&11.87  & 37.98  & 97.78 & 102.18 & 114.05 & 26.31 & 73.47 \\
Kimi-Audio-Base    & 9.7B                & 7.12  & 3.45 &22.35 & 127.07 & 8.55  & 70.30  & 88.11  & 96.11 & 51.26 \\
Step2-Audio-Mini-Base & 8.3B & 8.92 & 3.57&11.69 & 42.26 & 93.24 &131.80& 101.93& 42.09 & 123.17 \\ \bottomrule[1.3pt]
\end{tabular}
}
\caption{\label{lang-ood}Performance of ASR Models on Common Voice21 test set. Columns 3 to 11 in the first row are language codes according to ISO 639.}
\end{table*}

\begin{table}[!t]
\centering
\resizebox{\linewidth}{!}{
\begin{tabular}{lccccc}
\toprule[1.3pt]
\multirow{2}{*}{\textbf{Models}}                    & \multicolumn{3}{c}{\textbf{En}}            & \multicolumn{2}{c}{\textbf{Zh}}          \\ \cmidrule(lr){2-4} \cmidrule(lr){5-6}
 & AMI   & Earnings22 & Covo-DA & Covo-DA & MDT \\ \cmidrule(lr){1-6}
 SenseVoice               & 23.02 & 32.24      & 49.40        &    21.39       &15.03                   \\
SeamlessM4T                    & 1055  & 270.94     & 57.82        &   16.37        &  13.66                 \\ 
Kimi-Audio-Base                     &   1100    & 29.98           & 40.50        &    37.07       &  \textbf{7.47}                \\ \cmidrule(lr){1-6} 
Whisper-Large-V3               & 36.24 & 35.92      & 28.47        &     32.60      &  27.37               \\ 
+ \textsc{FT}, 1e-7 & 33.61 & 32.38 & 21.39 & 21.30 & 19.75\\
+ \textsc{DA} & 21.14 & 28.58 & 20.55 &16.03 & 17.28\\  \cmidrule(lr){1-6} 
Whisper-Turbo         & 38.59 & 36.90      & 25.12        &      34.56     & 26.38                   \\ 
+ \textsc{FT}, 1e-6 &  33.34 & 31.60 & 21.47 & 20.09 & 25.09 \\
+ \textsc{DA} & 19.10 & \textbf{17.33} & \textbf{18.83} & 15.93 & 14.11 \\ \cmidrule(lr){1-6} 
Qwen2-Audio-Base                    & 45.80 & 37.25      & 75.39        &        48.37   &       10.87           \\
+ \textsc{FT}, 1e-5 & 34.76&43.72&56.61 & 17.38 & 10.55  \\
+ \textsc{DA} & \textbf{15.53} & 24.10 & 19.10 & \textbf{12.73} & 7.49 \\ \bottomrule[1.3pt]

\end{tabular} 
}
\caption{\label{domain-ood}Performance of ASR Models on English and Chinese domain specific datasets. \textbf{FT} and \textbf{DA} refers to Fine-Tune and Data Augmentation, respectively.}
\end{table}

\begin{figure*}[t]
    \centering
    \scalebox{0.9}{
    \begin{subfigure}{0.49\textwidth}
        \includegraphics[width=\linewidth]{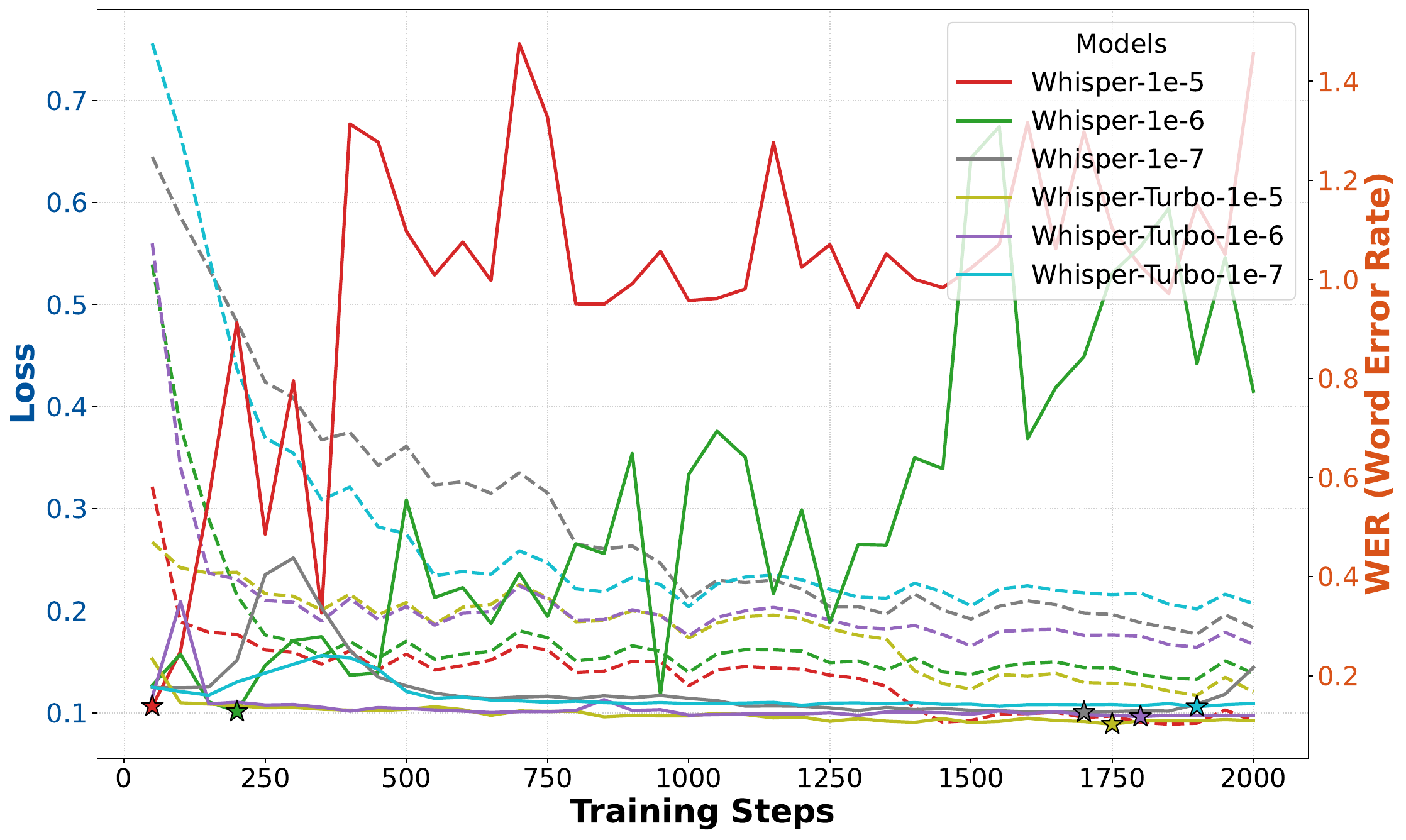} 
        \caption{Fine-tune Whisper/Whisper-Turbo on Chinese Corpus.}
        \label{fig:3x2_sub1}
    \end{subfigure}
    \hfill 
    \begin{subfigure}{0.49\textwidth}
        \includegraphics[width=\linewidth]{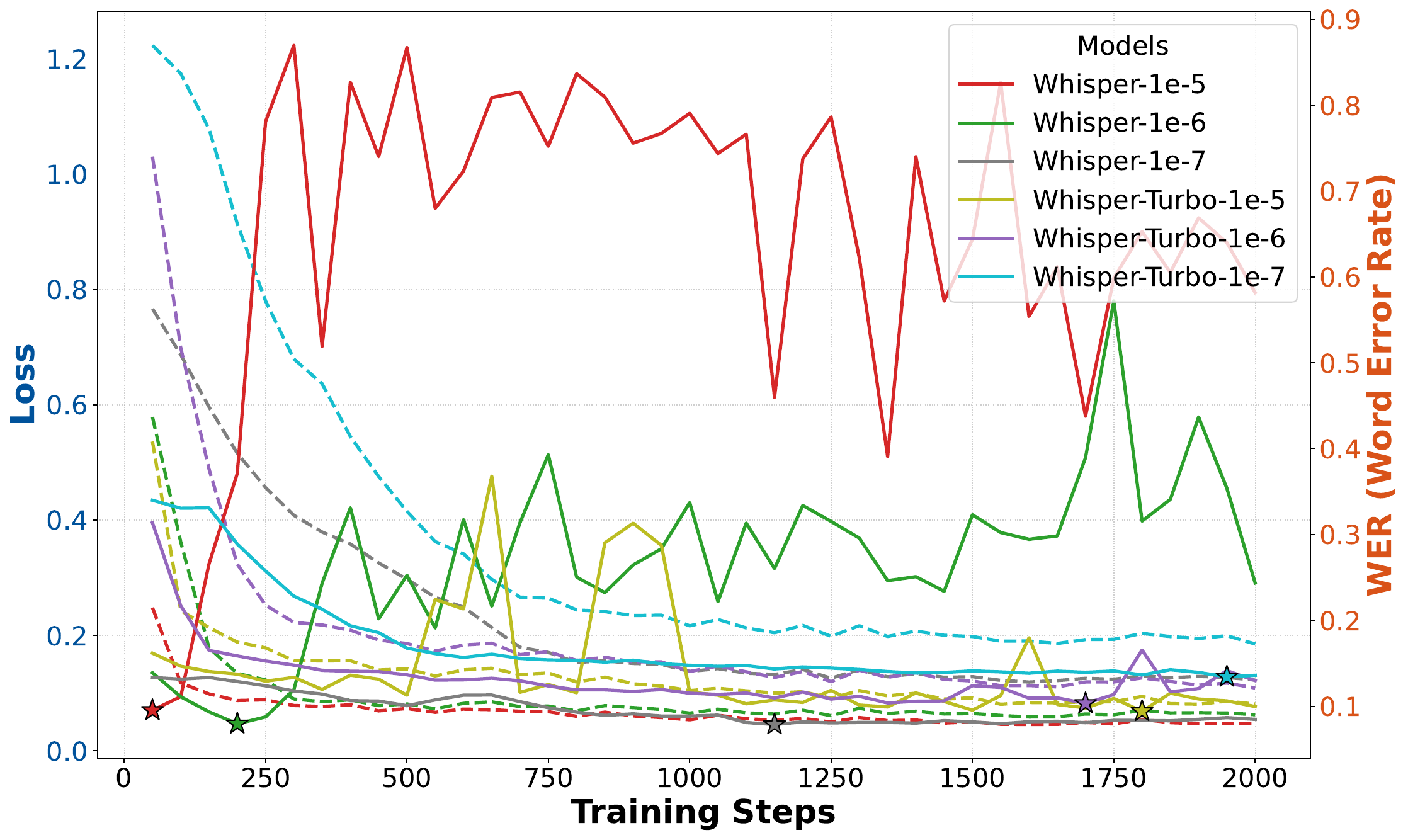}
        \caption{Fine-tune Whisper/Whisper-Turbo on Thai Corpus.}
        \label{fig:3x2_sub2}
    \end{subfigure}
    }
    
    \vspace{1em}

    \scalebox{0.9}{
    \begin{subfigure}{0.49\textwidth}
        \includegraphics[width=\linewidth]{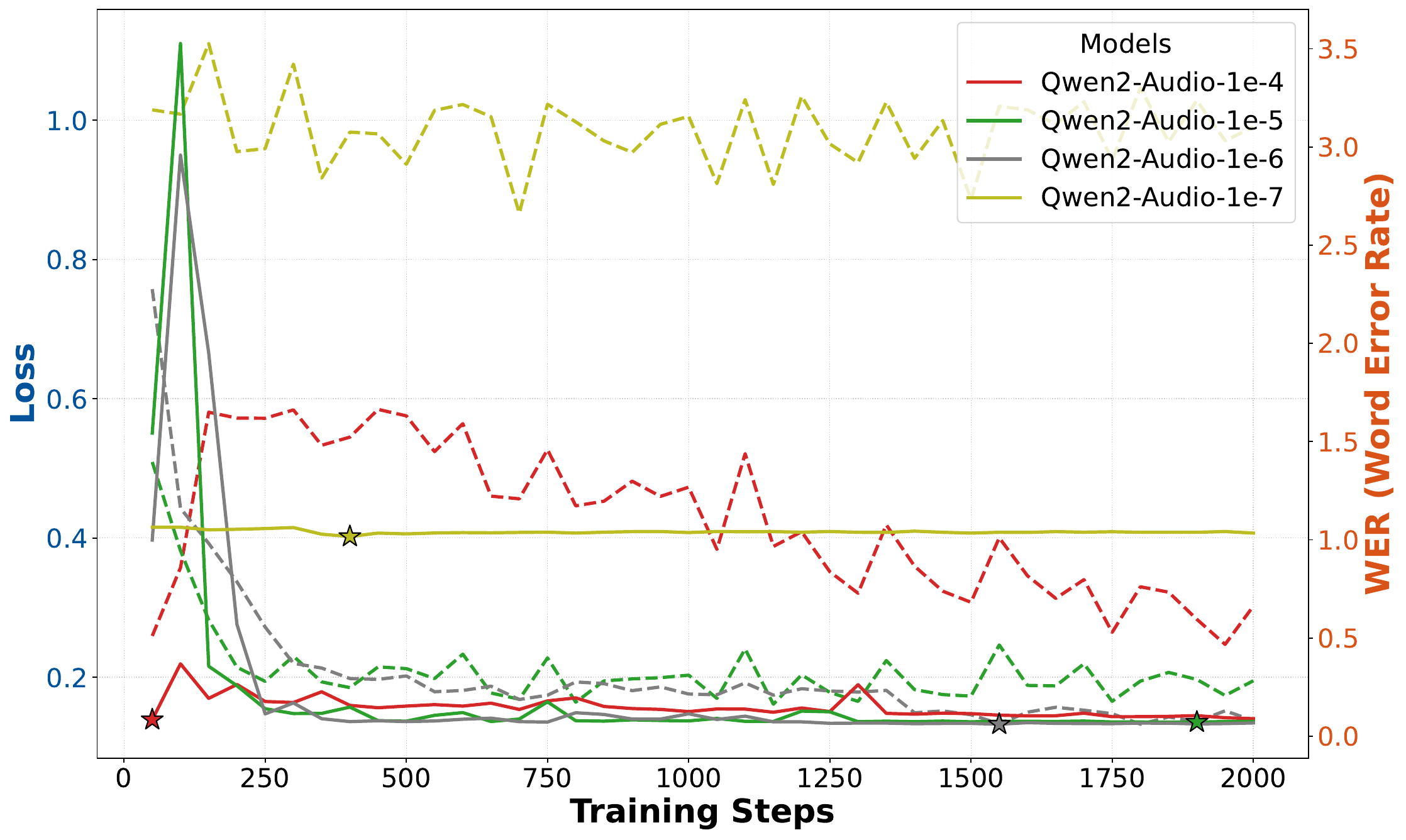}
        \caption{Fine-tune Qwen2-Audio-Base on Chinese Corpus.}
        \label{fig:3x2_sub3}
    \end{subfigure}
    \hfill
    \begin{subfigure}{0.49\textwidth}
        \includegraphics[width=\linewidth]{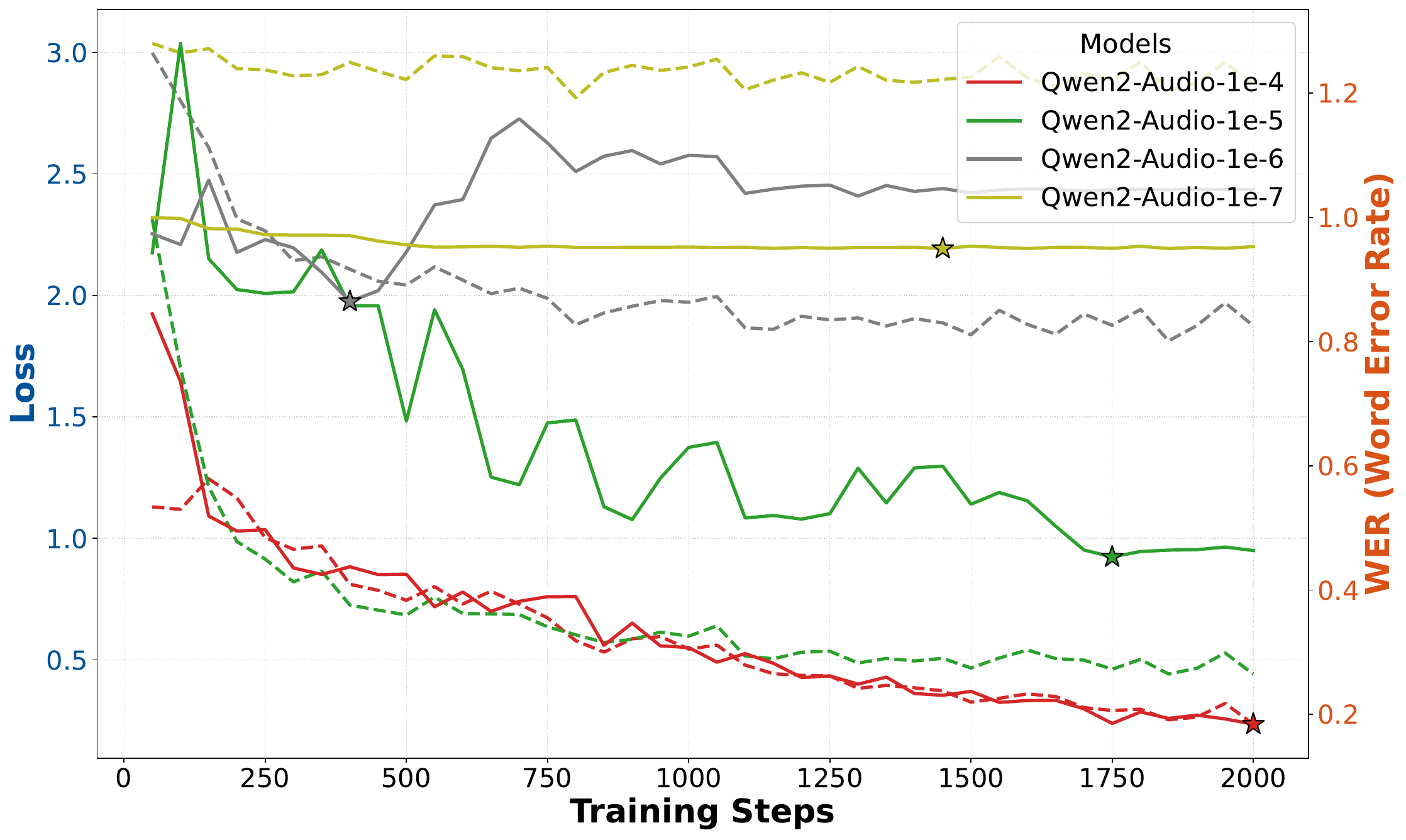}
        \caption{Fine-tune Qwen2-Audio-Base on Thai Corpus.}
        \label{fig:3x2_sub4}
    \end{subfigure}
    }
    \caption{Comparison of WER and Loss for different models fine-tuned on Chinese and Thai. The metrics are plotted against the number of training steps. In each plot, solid lines denote the WER, while dashed lines represent the training loss. The legend in the top-right corner specifies the model name and the learning rate used for its fine-tuning. The star marker indicates the lowest WER score.}
    \label{fig:all}
\end{figure*}

\section{Experiments}

\subsection{Models}
\label{models}
We fine-tune and evaluate a selection of representative state-of-the-art (SOTA) ASR models, including Whisper-Large-V3~\cite{radford2022robust_whisper}, its light-weight variant Whisper-Large-V3-Turbo, Qwen2-Audio~\cite{chu2024qwen2audiotechnicalreport}, and Kimi-Audio~\cite{kimiteam2025kimiaudiotechnicalreport}. For comprehensive comparison and to establish strong baselines, we also include SenseVoice~\cite{an2024funaudiollmvoiceunderstandinggeneration},  SeamlessM4T~\cite{DBLP:journals/corr/abs-2308-11596}, FireRedASR-AED~\cite{DBLP:journals/corr/abs-2501-14350}, and Step2-Audio~\cite{wu2025step}. To focus on the challenges of OOD speech recognition, we have excluded leading ASR models with limited language support, such as Canary~\cite{DBLP:conf/interspeech/PuvvadaZ0HKDMRC24} and Paraformer~\cite{DBLP:conf/interspeech/GaoZ0Y22}, from our experiments.

\subsection{Datasets}
We use a diverse suite of public speech datasets across two dimensions in our experiments:
(1) \textbf{Multilingual} datasets, include Fleurs~\cite{DBLP:conf/slt/ConneauMKZADRRB22}, Common Voice21~\cite{commonvoice:2020}, and GigaSpeech-2~\cite{yang2024gigaspeech}, which cover a wide range of languages and accents. (2) \textbf{Multi-Domain} datasets, includes AMI~\cite{DBLP:conf/mlmi/CarlettaABFGHKKKKLLLMPRW05} (English meetings), Earnings-22~\cite{DBLP:journals/corr/abs-2203-15591} (English financial calls), WeNet Speech~\cite{DBLP:conf/icassp/ZhangLGSYXXBCZW22} (YouTube), and MDT-AC001~\footnote{https://www.magicdatatech.cn/datasets/asr/mdt-asr-c001-mandarin-chinese-speech-recognition-corpus} (Chinese in-car speech).

Besides public datasets, we perform data augmentation on the CommonVoice corpus to create a new dataset, which we term \textbf{Covo-DA}. This process aims to simulate telephony channel conditions by applying a series of audio transformations including reducing the sampling rate and bit depth, applying band-pass filtering, and introducing distortion, saturation, and white noise.

\subsection{Experimental Setup}
For evaluation, we adhere to the official implementations and recommended settings provided by the respective model developers. For fine-tuning, we employ the AdamW optimizer with $\beta_1=0.9$ and $\beta_2=0.995$. The training is configured with a per-device batch size of 4 and 8 gradient accumulation steps. All fine-tuning experiments are conducted on a server equipped with 4$\times$NVIDIA H100 GPUs, resulting in a global batch size of 128. The learning rate boundaries $[\eta_{\min}, \eta_{\max}]$ are set to $[10^{-7}, 10^{-5}]$ for Whisper models and $[10^{-6}, 10^{-4}]$ for LLM-based ASR models, respectively. For our adaptive scheduler, the reference standard deviation $\sigma_{\text{ref}}$ is fixed at $0.5$.

\section{Results and Analysis}
Our experiments highlight the importance of careful data selection and hyper-parameter tuning in achieving robust domain adaptation. The framework is effective across both traditional and LLM-based ASR models, demonstrating its practical utility in real-world scenarios.

\subsection{Language and Domain OOD problems}
We evaluated the performance of ASR models on the CommonVoice21 test set. The results, presented in Tables~\ref{lang-ood} and~\ref{domain-ood}, reveal that nearly all models face significant OOD challenge in both language and domain. A clear performance gap is observed between high-resource languages, such as English (\textbf{En}) and Spanish (\textbf{Es}), and low-resource languages, where models perform poorly on Japanese (\textbf{Ja}), Thai (\textbf{Th}), and Vietnamese (\textbf{Vi}). Furthermore, all models struggle to generalize to domain-specific datasets, especially those with high levels of background noise, such as \textbf{phone calls} and \textbf{meeting recordings}.

\subsection{Hyper-Parameter Selection of Fine-tuning ASR Models}
We conduct fine-tuning on the three models detailed in Section 3.1, employing a range of hyperparameter combinations for Chinese and Thai. The training corpora consist of a 1000-hour subset of CommonVoice21-Zh combined with WeNetSpeech for Chinese, and a 1000-hour subset of CommonVoice21-Th with GigaSpeech2-Th for Thai. During training, both the loss and WER are recorded at intervals of 50 steps. The comprehensive results are presented in Figure~\ref{fig:all}.

Our key findings are summarized as follows:

(1) The relationship between training loss and WER is non-linear. As illustrated in Figures~\ref{fig:3x2_sub1} and~\ref{fig:3x2_sub2}, the Whisper model exhibits a monotonically decreasing loss, whereas its WER score fluctuates significantly. Notably, the optimal WER is often achieved in the early stages of training (e.g., around 1000 steps). This finding highlights the critical need for \textbf{well-designed early stopping criteria or pre-finetuning strategies to capture the best-performing checkpoint and prevent overfitting}.

(2) ASR models demonstrate \textbf{significant sensitivity to the learning rate, which is tied to both architecture and initial performance}. For instance, a learning rate of $10^{-6}$ proves insufficient for Whisper-Large-V3 to reduce WER on the Chinese task. In contrast, Whisper-Large-V3-Turbo, a variant with a shallower decoder, successfully converges with a higher learning rate of $10^{-5}$ on the same task. Similarly, while Qwen2-Audio achieves its optimal WER on Thai with a learning rate of $10^{-6}$, effective fine-tuning on Chinese necessitates an increase to $10^{-4}$.

We attribute the learning rate sensitivity in Whisper models primarily to their decoder depth. The 32-layer decoder of Whisper-Large-V3 requires a conservative learning rate to prevent training instability, which can manifest as out-of-vocabulary (OOV) token hallucination or task confusion (i.e., performing translation instead of transcription). In contrast, the much shallower 4-layer decoder in Whisper-Large-V3-Turbo exhibits greater stability, tolerating higher learning rates.

Conversely, LLM-based ASR models often benefit from larger learning rates. Their strong intrinsic language capabilities enable rapid convergence within a few hundred steps, particularly on languages where the model has strong baseline performance. For example, with a $10^{-4}$ learning rate, Qwen2-Audio-Base achieves a competitive WER on Chinese after only 200 fine-tuning steps.

\subsection{Data Augmentation for Domain Specific Fine-tuning}

We apply our data augmentation strategy across multiple scenario datasets in both Chinese and English. The Chinese training corpus is the same as described in Section 4.2, while the English training corpus is entirely sourced from CommonVoice21. The training volume remains at 1,000 hours. We fine-tune ASR models on English and Chinese, using learning rate in Section~4.2 and employing early stopping. As shown in Table~\ref{domain-ood}, the model's performance on both original and augmented data demonstrates the substantial benefits of our domain-specific augmentation. A clear performance hierarchy emerges: \textbf{fine-tuning on augmented data yields the lowest WER}, followed by fine-tuning on original data, with both methods significantly outperforming the baseline pre-trained model. This result validates the effectiveness of our automated augmentation pipeline and our empirical guidelines for selecting fine-tuning learning rates.

\section{Conclusion}
This paper presents a unified framework for the domain-specific fine-tuning of ASR models, designed to systematically address key challenges in adapting both traditional architectures and modern LLM-based systems. Our framework integrates two core components: a rigorous hyperparameter optimization process to maximize metric-driven performance, and a principled data augmentation strategy to enhance model generalization and robustness. Through empirical evaluations across a diverse range of models and datasets, we consistently demonstrate the effectiveness of this approach via significant improvements in recognition accuracy. Ultimately, this work provides practitioners with a clear and validated set of guidelines, aiming to transition the field from ad-hoc fine-tuning to a more structured and reliable paradigm for ASR domain adaptation.

\bibliographystyle{IEEEbib}
\bibliography{strings,refs}

\end{document}